\documentclass[a4paper]{siamart1116}
\usepackage{mathtools}
\usepackage{epstopdf}
\usepackage{algorithm,algorithmic}
\usepackage{amsmath}
\usepackage{amssymb}
\graphicspath{{images_color/}}
\usepackage{hyperref}
\usepackage[square,numbers]{natbib}
\usepackage{graphics}
\graphicspath{{figures/}}

\usepackage{subcaption}

\usepackage{enumitem}
\setlist[itemize,1]{leftmargin=\dimexpr 3em}

\usepackage{extdash}
\newcommand{\indententry}{
\hspace{1em}\ignorespaces
}

\title{The Chunks and Tasks Matrix Library 2.0
\thanks{\today
\funding{This work was supported by the Swedish national strategic e-science research program (eSSENCE).}}}

\author{Emanuel H. Rubensson%
\thanks{Division of Scientific Computing, Department of Information Technology, Uppsala University, Box 337, SE-751 05 Uppsala, Sweden (\email{emanuel.rubensson@it.uu.se}, \email{elias.rudberg@it.uu.se}, \email{nastja.kruchinina@gmail.com}, \email{anton.artemov@it.uu.se}).}
\and
Elias Rudberg%
\footnotemark[2]
\and
Anastasia Kruchinina%
\footnotemark[2]
\and
Anton G. Artemov%
\footnotemark[2]
}

\begin{document}
\maketitle

\begin{abstract}
  We present a C++ header-only parallel sparse matrix library, based
  on sparse quadtree representation of matrices using the Chunks and
  Tasks programming model. The library implements a number of sparse
  matrix algorithms for distributed memory parallelization that are
  able to dynamically exploit data locality to avoid movement of data.
  This is demonstrated for the example of block-sparse matrix-matrix
  multiplication applied to three sequences of matrices with different
  nonzero structure, using the CHT-MPI~2.0 runtime library
  implementation of the Chunks and Tasks model. The runtime library
  succeeds to dynamically load balance the calculation regardless of
  the sparsity structure.
\end{abstract}

\begin{keywords}
  Chunks and Tasks programming model, block-sparse, parallelization, quadtree, sparse matrices, task-based programming 
\end{keywords}

\section{Motivation and significance} \label{sec:motivation_and_significance}
Computing the product of two matrices is one of the most fundamental
operations in scientific computing~\cite{gao2020systematic}.  Matrix-matrix multiplication
is a core operation in for example combinatorics
\cite{suitsparse}, deep learning~\cite{chetlur2014cudnn}, and
electronic structure theory~\cite{Bowler_Miyazaki_review_2012,
  olivares_2010}. Other commonly used linear algebra
operations such as linear solves, matrix inversions and factorizations
are reduced to matrix-matrix multiplications in efficient blocked
implementations~\cite{elmroth_sirev}.

In the dense matrix case, efficient software typically makes
use of the Basic Linear Algebra Subprograms (BLAS). This makes the
code portable and gives high performance when linked to an optimized
BLAS library.  Parallelization schemes for dense linear algebra
problems are often based on static distribution of data, such as
the two-dimensional block and block-cyclic mappings of the matrix onto
a two-dimensional process grid, for example used in the SUMMA
algorithm for parallel matrix-matrix multiplication~\cite{summa}
and the HPL implementation of the HPLinpack benchmark~\cite{hpl_impl},
respectively.
Since the matrices are dense, all computation and communication costs
are known in advance and, in general, the same scheme can be used
regardless of the application since the only parameters that may vary
are the matrix dimensions.

Software which makes use of parallel sparse matrix-matrix multiplication,
on the other hand, often resorts to specialized implementations
relying on a priori information about the matrix sparsity pattern
for the application at hand. Such implementations can be found in a
number of electronic structure codes~\cite{onetep-sparsematrix, WeberEtAlMidpointMmul2015}.  Although great performance is
often achieved, the benefits of general purpose algorithms and
libraries are missing. 

A class of general purpose algorithms
for parallel sparse matrix-matrix multiplication
is based on random permutation
of rows and columns of the input matrices, with the purpose to evenly
spread out the nonzero entries over the
matrix~\cite{multiple_level_spmul, Borstnik2014, BulucGilbert2012}.
After the random
permutation some method for dense matrices is applied but with the
local block products replaced by sparse products. This gives an
approach that is generally applicable without need for a priori
information about the sparsity pattern, but the cost you have to pay
is that the nonzero structure and any data locality is destroyed.
This inability to exploit the structure of the input matrices has
serious consequences for the cost of communication~\cite{rubensson2016locality}.

Without a priori information about the sparsity pattern and without
compromising on performance, the mapping of work and data to physical
resources has to be done dynamically during the calculation. The
development of such schemes directly on top of the standard Message
Passing Interface (MPI) is an overwhelming task. This is a reason
why one has been unable to combine performance and portability of
algorithms and software.  The use of some task-parallel framework
seems like a viable approach.  However, most task-parallel frameworks
are primarily developed for shared memory and typically resort to
static distribution or centralized management of data when extended to
computer clusters with distributed memory.

We present here the Chunks and Tasks Matrix Library which can be used
to implement algorithms that do not make assumptions about the
structure of input matrices, but efficiently exploits data locality
when available. Critical to the development of this library is its use
of the Chunks and Tasks programming model~\cite{CHT-PARCO-2014},
which provides abstractions
for both work and data, and therefore makes the implementation of
locality-aware parallel sparse matrix algorithms feasible. The library
is written in C++ and may be used either to parallelize performance
critical matrix operations of a C++ program or as part of a
full-fledged Chunks and Tasks application.

The Chunks and Tasks Matrix Library has been indispensable in the
development, implementation, and analysis of a number of novel
parallel sparse matrix algorithms for distributed memory systems,
including a general purpose parallel sparse matrix-matrix multiply
efficiently exploiting locality of nonzero matrix entries
\cite{rubensson2016locality}, a sparse approximate
matrix-matrix multiply for matrices with decay
\cite{2019arXiv190608148A,2020arXiv200510680A}, localized inverse
factorization
\cite{Rubensson2018localized,Artemov2018parallelization}, a new
communication-avoiding divide and conquer method for inverse
factorization of symmetric positive definite matrices, and density
matrix purification \cite{2019arXiv190912533K}.

\section{Software description} \label{sec:software_description}
The Chunks and Tasks Matrix Library is a C++ header-only library,
implemented using the Chunks and Tasks programming model. To use the
library in your C++ program, you need a Chunks and Tasks runtime
library and an implementation of the Basic Linear Algebra Subprograms
(BLAS)~\cite{blas3}. Open source Chunks and Tasks library implementations are
publicly available at \verb|chunks-and-tasks.org|.

In the Chunks and Tasks Matrix Library, matrices are represented using
sparse quaternary trees (quadtrees)~\cite{quadtreeWise1984}.
In this representation a matrix
is either 1) identically zero, 2) stored using a data structure for
small matrices, or 3) split into four quadrants, each a submatrix
recursively represented by a sparse quadtree. Matrix operations are
formulated as recursive algorithms traversing nonzero branches of the
matrix quadtrees. Quadtrees provide an efficient way to squeeze out
zero entries from the representation and to expose parallelism in both
data and work.

\subsection{Software Architecture}
The main components of the library are the following:
\begin{itemize}
\item a matrix chunk template implementing the sparse quadtree
  representation of matrices using the Chunks and Tasks model,
\item a number of task templates implementing various recursive
  algorithms operating on sparse quadtrees using the Chunks and Tasks
  model,
\item and three stand-alone matrix libraries for leaf matrix
  representation.
\end{itemize}
The chunk template for quadtree matrix representation is parameterized
with a leaf matrix type, used for matrix representation at leaf nodes.
This leaf matrix type may or may not be one of the types distributed
with the Chunks and Tasks Matrix Library and may use a sparse or dense
representation.
A non-leaf node/chunk in the quadtree is the parent of four
chunks. Each child chunk is referred to by its chunk identifier and
represents a quadrant of the matrix corresponding to that non-leaf
node. A Chunks and Tasks nil identifier is used to refer to
submatrices that are identically zero.

The task templates for matrix operations are also parameterized with
the leaf matrix type. The key component in each task implementation is
an {\texttt{execute}} function called by the runtime library to carry
out task execution. This function performs operations on a single
level in the quadtree. For non-leaf input the {\texttt{execute}}
function typically registers tasks operating on children of the input
chunks. At the leaf level, leaf matrix functionality corresponding to
the given task type is invoked. Matrix sparsity, i.e. zero branches in
the quadtree, is handled in a fallback {\texttt{execute}} function,
called by the Chunks and Tasks runtime library whenever one of the
input matrix chunk identifiers is nil.

Three stand-alone leaf matrix types are distributed with the library,
each in a subdirectory with the following names:
\begin{itemize}
\item \verb|basic_matrix_lib :| Dense matrix representation using a
  standard column-wise layout of matrix elements~\cite{CHT-PARCO-2014}.
\item \verb|block_sparse_matrix_lib :| Block-sparse matrix
  representation where submatrices (blocks) of uniform size are laid
  out in a twodimensional array~\cite{rubensson2016locality}. Zero
  submatrices are neither stored
  nor referenced.
\item \verb|hierarchical_block_sparse_lib :| Sparse quadtree matrix
  representation resembling the implementation at the Chunks and Tasks
  level~\cite{2019arXiv190608148A}.
\end{itemize}

The quadtree chunk template describes a data structure suitable for
parallel distribution of matrices, but does not specify where data
should be stored. The task templates for matrix operations describe
recursive algorithms suitable for parallel execution, but do not
make task scheduling decisions. The leaf matrix libraries distributed
with the library are serial, unless linked to a threaded
implementation of BLAS. Parallelism is achieved when the library is
used together with a parallel implementation of the Chunks and Tasks
model.

\subsection{Software Functionalities}
The main functionality provided by the task templates in the Chunks
and Tasks Matrix Library, in combination with the leaf matrix types,
can be categorized as follows:
\begin{itemize}
\item \emph{Assignment from and extraction of matrix elements}: task types
  to construct a matrix from vectors with row and column indices and
  values and task types to extract elements of a matrix specified by
  vectors with row and column indices.
\item \emph{Matrix addition}: addition task types including regular matrix
  addition and addition of a matrix with a scaled identity matrix.
\item \emph{Matrix-matrix multiplication}: multiplication task types including
  regular and symmetric multiplication, symmetric matrix square, and
  symmetric rank-k construction, as well as sparse approximate
   multiplication.
\item \emph{Inverse factorization}: task types for inverse factorization of
  symmetric positive definite matrices including inverse Cholesky and
  localized inverse factorization.
\item \emph{Truncation}: task types for removal of small matrix elements
  with different variants of error control.
\end{itemize}
All task types are able to handle and exploit matrix sparsity. 
This list of functionality reflects the fact that the main motivation
for the development of the library has been its application to
large-scale electronic structure calculations.
The library also includes a number of auxiliary task templates not listed here. 

\section{Illustrative Examples} \label{sec:illustrative_examples}
We illustrate in this section the capabilities of the library as
outlined earlier in this article, using the general sparse
matrix-matrix multiplication as an example.  We
construct sequences of matrices of growing size and, for each matrix
in each sequence, compute the product of the matrix with itself. For
each sequence, the setup is such that the number of floating point
operations needed for the products is proportional to the matrix
dimension. Scaling up the computing resources with the matrix
dimension gives a weak scaling experiment for each sequence.
Three matrix sequences are constructed as follows:
\begin{itemize}
\item \emph{Banded}: banded matrices with bandwidth $2\times 3000 +1$.
\item \emph{Growing block}: banded matrices, again with bandwidth
   $2\times 3000 +1$, but with a single large dense block added to the
  upper left corner. The size of the block is chosen so that the total
  number of floating point multiplications for the matrix multiply is
  doubled compared to the banded matrix. This gives a block size  
  growing with matrix size.
\item \emph{Random blocks}: banded matrices, again with bandwidth
  $2\times 3000 +1$, but with a number of dense equally sized blocks
  placed at random positions along the main diagonal, without
  overlap. The number of blocks is proportional to the matrix
  dimension. The size of the blocks is chosen so that the total number
  of floating point multiplications for the matrix multiply is doubled
  compared to the banded matrix. The block size stays
  essentially constant as the matrix size increases.
\end{itemize}
The same set of matrix dimensions is used in all three experiments.
The experiment setup with matrix sizes, block sizes, etc is given in
Table~\ref{table1}.

\begin{table}[]
\centering
\resizebox{\columnwidth}{!}{%
\begin{tabular}{lccccccc}
\multicolumn{8}{c}{\textbf{Weak scaling experiment setup}} \\ 
\hline
Matrix Size      &  $10^5$     & $2 \times 10^5$      & $4 \times 10^5$      & $8 \times 10^5$      & $1.6 \times 10^6$ & $3.2 \times 10^6$    & $6.4 \times 10^6$    \\ 
No.\ of worker processes & 2    & 4    & 8    & 16    & 32    & 64 & 128   \\ \hline
\multicolumn{8}{l}{\textbf{Banded}} \\ 
\indententry Tflop & 7.022 & 14.22 & 28.63 & 57.44 & 115.1 & 230.3 & 460.8\\ \hline
\multicolumn{8}{l}{\textbf{Growing block}} \\ 
\indententry Size of block & 15716    & 19652    & 24621    & 30899    & 38825    & 48828    & 61446    \\ 
\indententry Tflop            & $14.04$ & $28.45$ & $57.26$ & $114.9$ & $230.1$ & $460.6$ & $921.6$ \\ \hline
\multicolumn{8}{l}{\textbf{Random blocks}} \\ 
\indententry Size of blocks & 15716    & 15705    & 15700    & 15697    & 15696    & 15695    & 15695    \\ 
\indententry No.\ of blocks & 1    & 2    & 4    & 8    & 16    & 32    & 64    \\ 
\indententry Tflop            & $14.04$ & $28.45$ & $57.26$ & $114.9$ & $230.1$ & $460.6$ & $921.6$ \\ \hline
\end{tabular}
}
\caption{Parameters used to set up the weak scaling experiments as
  described in the text. The Tflop values are the numbers of floating
  point operations needed for each sparse matrix-matrix multiplication
  measured in teraflops.}
\label{table1}
\end{table}

All experiments were performed on the Beskow cluster at the PDC Center
for High Performance Computing at the KTH Royal Institute of
Technology in Stockholm, using the soon publicly available Chunks and Tasks
library CHT-MPI~2.0\footnote{The release of CHT-MPI~2.0 is under preparation.}.  Beskow is a Cray XC40 system with 2060 compute
nodes based on Intel Haswell and Broadwell processors and a Cray Aries
interconnect with Dragonfly topology.  The present study made use of
the Haswell nodes, each equipped with two 16-core Intel Xeon E5-2698v3
2.3 GHz CPUs and 64 gigabytes of memory. Each compute node has a
theoretical peak performance of approximately 1280
Gflop/s~\cite{2019arXiv190608148A}. The CHT-MPI~2.0 library uses the
Message Passing Interface (MPI) for communication between nodes.  The
code was compiled with GCC 9.3.0 and Cray MPICH 7.7.14.
In CHT-MPI~2.0 an MPI parent process runs the main program and spawns
a number of MPI worker processes that execute tasks. The program was
configured to use 1 worker process per node and 31 worker threads per
worker process, executing tasks. Thus, one core per node is left for
threads handling communication with other nodes. The chunk cache size
for each CHT-MPI~2.0 worker process was set to 4~GB.

The leaf matrix dimension was set to $2048 \times 2048$. Leaf matrices were
represented using the block-sparse matrix library with leaf internal
blocksize $64 \times 64$ and matrix elements were stored in double
precision. Single threaded OpenBLAS~0.3.10 \cite{OpenBLAS} was used for BLAS
operations on submatrices within the block-sparse leaf matrix library.
The input matrices were constructed distributed over the worker
processes.

\begin{figure*}[ht]
  \centering
  \captionsetup{subrefformat=parens}
  \begin{subfigure}[b]{0.48\linewidth}
    \includegraphics[width=1\linewidth]{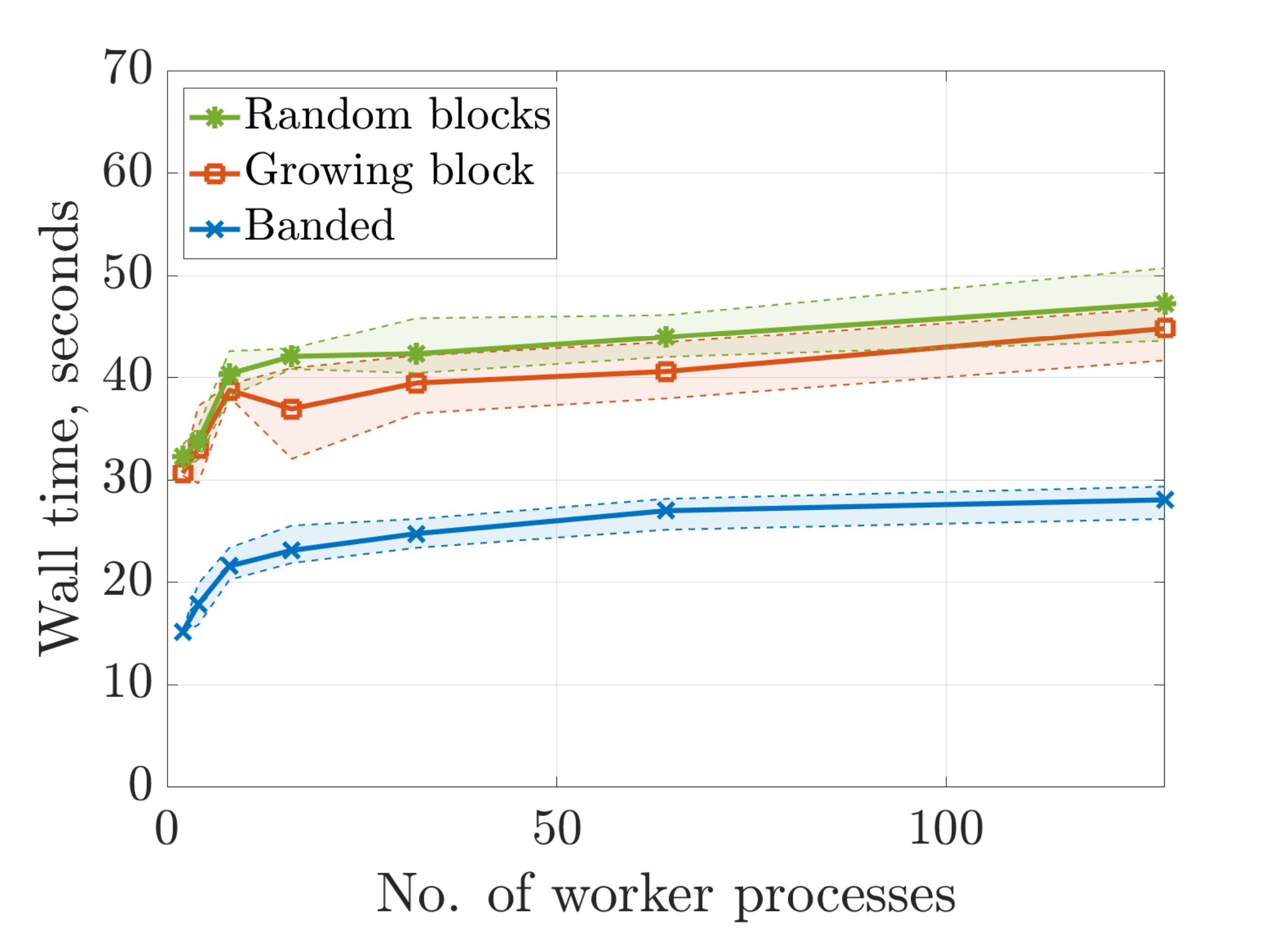}
    \caption{Weak scaling\label{panel_time}}
  \end{subfigure}
  \begin{subfigure}[b]{0.48\linewidth}
    \includegraphics[width=1\linewidth]{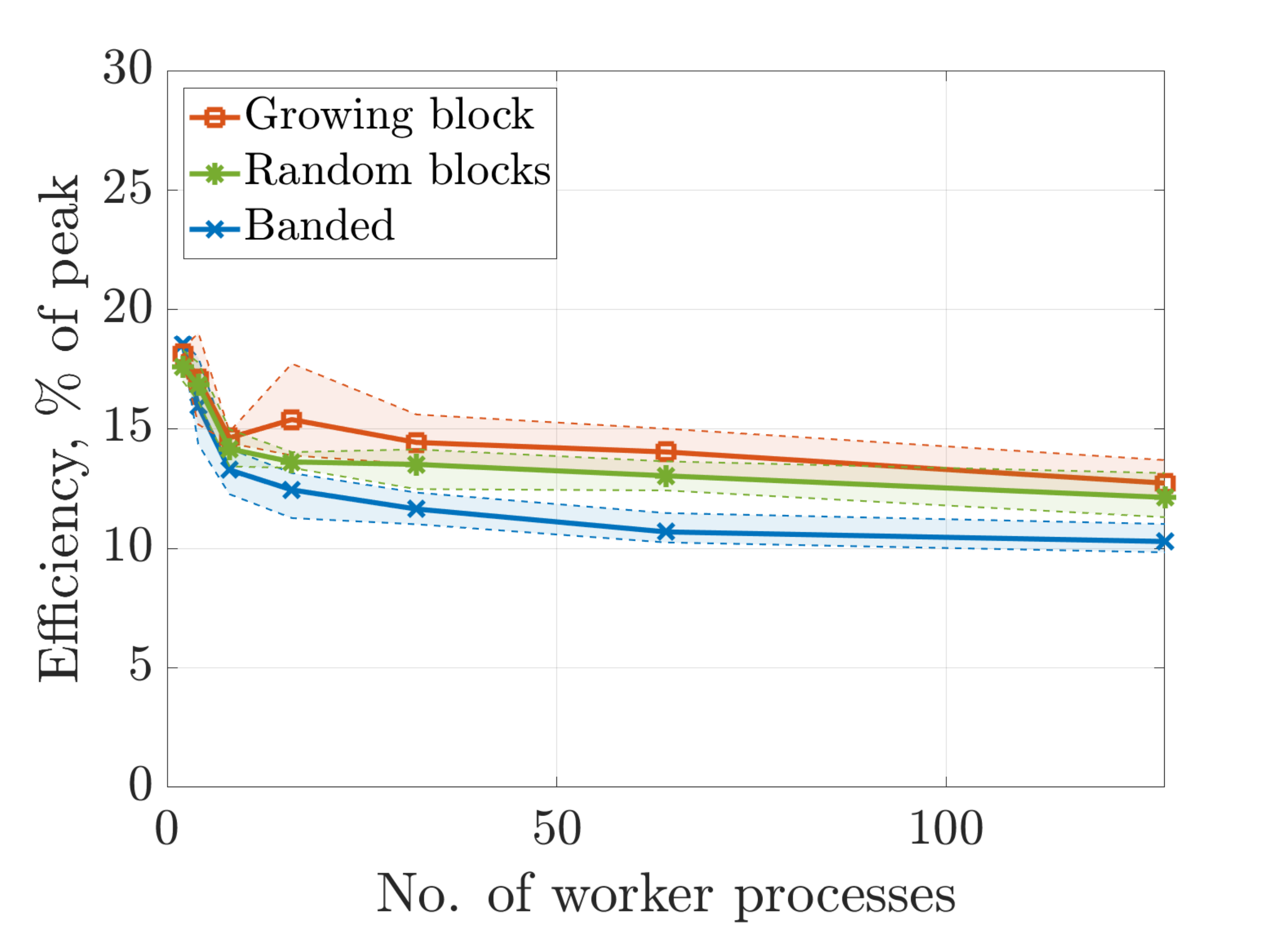}
    \caption{Efficiency\label{panel_eff}}
  \end{subfigure}
    \begin{subfigure}[b]{0.48\linewidth}
      \includegraphics[width=1\linewidth]{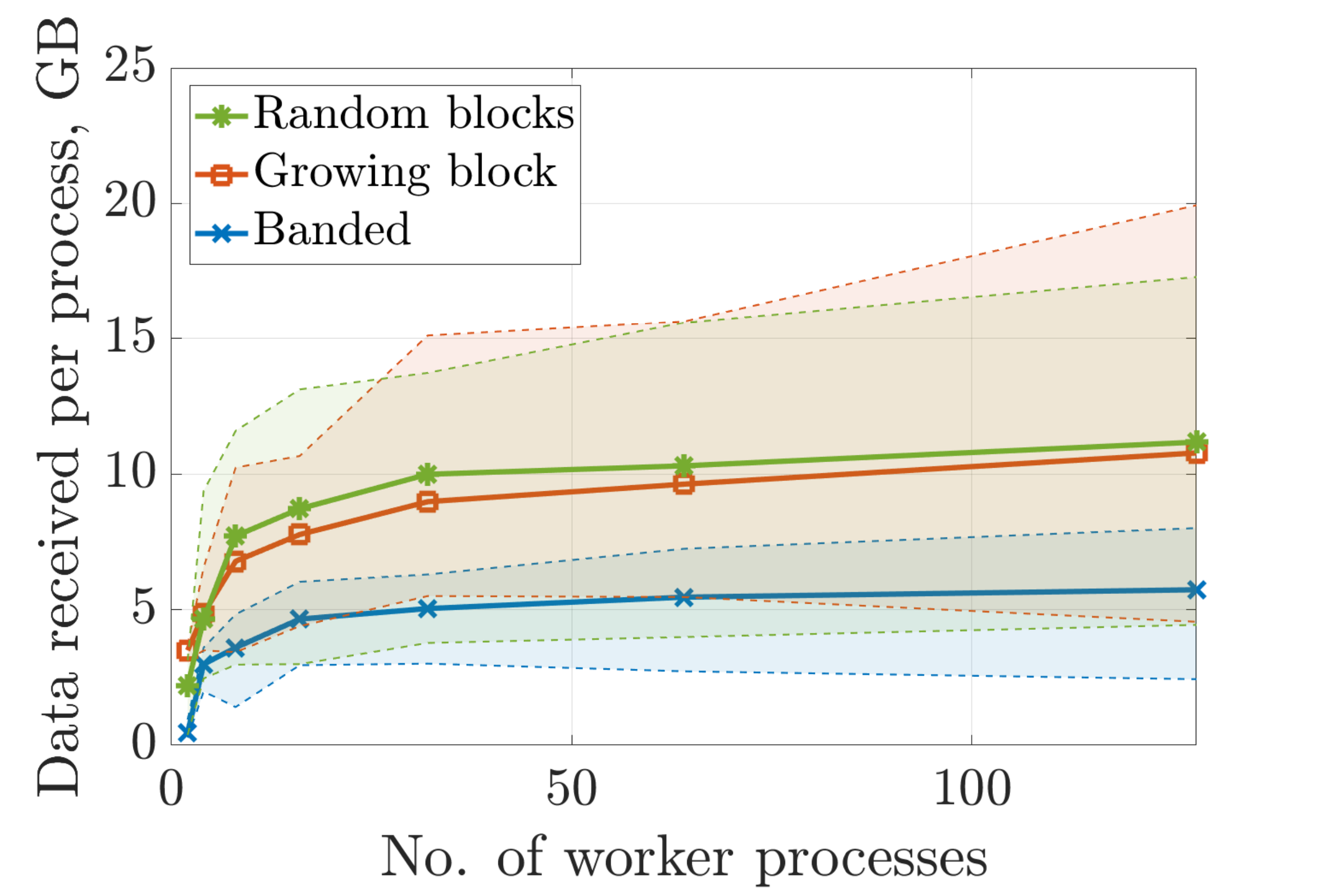}
    \caption{Communication\label{panel_data}}
    \end{subfigure}
  \caption{Performance of sparse matrix-matrix multiplication using
    the Chunks and Tasks Matrix Library, demonstrated for three
    different weak scaling test cases described in the
    text. Panel~\subref{panel_time}: Wall times.
    Panel~\subref{panel_eff}: Efficiency measured as the ratio between
    the achieved floating point operations per second and the
    theoretical peak performance of the compute nodes employed.
    Panel~\subref{panel_data}: Data received per worker process.  In
    both Panels~\subref{panel_time} and~\subref{panel_eff}, the
    solid lines show the average values over four test runs while the
    dashed lines show the minimum and maximum values. In
    Panel~\subref{panel_data}, the average, minimum and maximum values
    are taken not only over the four test runs but also over all
    worker processes.}
\label{fig:timing_and_efficiency}
\end{figure*}

Figure~\ref{panel_time} shows the wall time as a function of number of
worker processes for the three test cases. The figure shows
the average, minimum, and maximum wall times of 4 repeated test
runs. The experiment confirms previous theoretical and experimental
results for the banded matrix, showing a wall time increasing
logarithmically with the number of nodes~\cite{rubensson2016locality}.
We can conclude that the library succeeds in dynamically balancing the
load also in the other two cases, where the nonzero structure is not
uniform. Although the number of scalar operations is twice that
of the banded matrix test case, the wall times are considerably less
than twice the wall times of the banded matrix case. One possible
explanation is that the large added blocks give computations with
larger arithmetic intensity, i.e.\ with more floating point
operations per nonzero element.  Tasks operating on the same chunk are
likely to be executed by the same worker process. Note that
CHT-MPI~2.0 makes use of work stealing between worker processes with
stolen tasks chosen from the tree of tasks based on a breadth first
strategy. The higher
efficiency for the ``Growing block'' and ``Random blocks'' test cases
can also be seen in Figure~\ref{panel_eff} which shows the efficiency
measured as the ratio between the number of floating point
operations per second and the theoretical peak performance of the
employed compute nodes.
Note that the matrix library dynamically balances the workload while
exploiting locality in the nonzero structure to reduce communication.
The library makes no use of a priori information regarding the nonzero
structure of the matrices.

Figure~\ref{panel_data} shows the communication measured as the amount
of data received per process during the multiply. The figure shows the
average over all processes and the four repeated test runs. The
maximum (minimum) values are taken as the maximum (minimum) over all
processes and the four repeated test runs. As a reference, we mention that
the largest
matrix in the banded test case has about 3.8 billion nonzero elements
corresponding to 307~GB of memory in double precision floating point
representation.

\section{Impact}
The Chunks and Tasks Matrix Library has enabled the development of a
number of sparse matrix algorithms for distributed memory
parallelization, as discussed in
Section~\ref{sec:motivation_and_significance} and as further described
in Section~\ref{sec:software_description}. These algorithms have
in common their ability to dynamically exploit data locality to avoid
movement of data, as demonstrated in
Section~\ref{sec:illustrative_examples} for sparse
matrix-matrix multiplication. They also have in common that they are the
most performance critical sparse matrix operations in large-scale
electronic structure calculations. In particular,
different variants of matrix-matrix
multiplication have received ample attention among developers of
methods and software for electronic structure calculations. The Chunks
and Tasks Matrix Library implements task types for parallel sparse
matrix-matrix multiplication that combine performance and portability in a
unique way as described in Section~\ref{sec:motivation_and_significance}.
Yet the most important
contribution of the Chunks and Tasks Matrix Library may be that it is
showcasing new technology making the achievements above possible.

The Chunks and Tasks Matrix Library represents a paradigm shift for
the parallelization of sparse matrix operations. This shift means a
change in the division of responsibilities between the developer of an
application and the runtime library. For any parallel programming
model this division of responsibilities is basically defined by the
programming interface of the library or language associated with the
model. An application programmer using the Chunks and Tasks
programming model worries about exposing parallelism in data and work
to the runtime library but not about the mapping of this parallelism
to physical resources. This means that the application programmer does
not have to think explicitly about where data should be stored, which
process should execute a particular task, synchronization, or
communication. At the same time we have seen that the model is
expressive enough to describe complex parallel data structures and
algorithms.  The model imposes certain restrictions for access to data
in user code, making efficient and scalable runtime implementations
possible. In particular, the CHT-MPI~2.0 implementation uses a completely
decentralized management of both work and data, atypical to models
where the data distribution is handled by the runtime.  Those
properties of the Chunks and Tasks programming model are all inherited
by the Chunks and Tasks Matrix Library and makes the library a potential game
changer for the development and implementation of parallel sparse
matrix operations.

The Chunks and Tasks Matrix Library is released under the modified BSD
license and is publicly available for download at
\verb|chunks-and-tasks.org|.

\section{Conclusions}
We have presented the Chunks and Tasks Matrix Library which is a
sparse matrix library based on the Chunks and Tasks programming model
and sparse quadtree representation of matrices. An advantage of the
quadtree representation is that large zero blocks of a sparse matrix
can be skipped already at a high level, leading to improved
scalability in cases with data locality. The library provides a
convenient set of routines for handling large sparse matrices in a
distributed setting with main focus on multiplication-heavy
algorithms, which are common in electronic structure calculations.
Future work and extensions of the library include continued
development of algorithms for sparse matrix-matrix multiplication and
the combination of outer product based matrix-matrix multiplication
with the quadtree representation, to improve the scaling behavior in
cases with poor data locality.

\section*{Acknowledgements}

Computational resources were provided by the Swedish National Infrastructure for Computing (SNIC) at the PDC Center for High Performance Computing, KTH Royal Institute of Technology in Stockholm.

\bibliography{references} 

\end{document}